# Fabrication of Deep-Sub-Micrometer NbN/AlN/NbN Epitaxial Junctions on a Si Substrate

Wei Qiu, Hirotaka Terai

Advanced ICT Research Institute, National Institute of Information and Communications Technology, 588-2 Iwaoka, Nishi-ku, Kobe, 651-2492 Japan

E-mail: qiuwei@nict.go.jp

**Abstract**

We have developed a novel fabrication process for ultra-small, full-epitaxial NbN Josephson junctions on a silicon (Si) substrate. A full-epitaxial NbN/AlN/NbN tri-layer was grown on a Si (100) wafer with a (200)-oriented TiN buffer layer. It was patterned into Josephson junctions by an electron beam lithography (EBL) for junction definition followed by a reactive ion etch (RIE). A chemical mechanical polishing (CMP) process and an additional RIE by using $CHF_3$ gas formed reliable electrical contacts between the junction counter electrodes and the wiring layer. All fabricated junctions, with a junction size down to 0.27 μm in diameter, showed excellent current-voltage characteristics with a clear gap structure and a small sub-gap leakage current. The dielectric layer of $SiO_2$ that served as an insulator between base and counter electrodes was removed in a wet etching process using a buffered HF solution. We have confirmed that the quality of the junctions was maintained after the removal of the $SiO_2$ dielectric layer.

1. Introduction

Nitride superconductors, including niobium nitride (NbN), titanium nitride (TiN), and niobium titanium nitride (NbTiN), have been used in a wide variety of applications, due to the high transition temperature, $T_c$, and large superconducting energy gap, $2\Delta$. Applications include superconductor-insulator-superconductor (SIS) mixers [1], superconducting single-flux quantum circuits (SFQ) [2], superconducting nanowire single-photon detector (SNSPD) [3], and microwave kinetic inductance detectors (MIDs) [4], We have developed NbN/AlN/NbN epitaxial junctions for superconducting quantum bit (qubit) as an alternative technology for Al/AlO$_x$/Al junctions [5-7]. The Al/AlO$_x$/Al junction has been used for most studies of superconducting qubit systems to date [8-11] because the AlO$_x$ tunnel barrier is a stable passivity and ultra-small Al-junctions can be easily fabricated by an electron beam lithography (EBL) and an angle evaporation. However, the amorphous AlO$_x$ tunnel barrier fabricated by the oxidization of an evaporated Al layer contains microscopic two-level systems (TLSs) that are well-known decoherence sources in superconducting qubits. Among many efforts to eliminate the impact of TLSs on qubit coherence time, developing full-epitaxial tunnel junction is one of the important approaches [12]. Full-epitaxial NbN/AlN/NbN tunnel junction would provide a similar solution to that extend, where the AlN tunnel barrier can grow in a cubic structure instead of a stable hexagonal bulk phase with the piezoelectricity [13]. We successfully observed clear Rabi oscillation in the transmon qubit consisting of epitaxial NbN/AlN/NbN junctions. However, the obtained energy relaxation time $T_1$ was about 500 ns due to a large dielectric loss from the MgO substrate and the MgO/NbN interface [6]. To reduce the substrate related loss, we have developed the growth technique of a full-epitaxial NbN/AlN/NbN tri-layer on a single-crystal Si (100) wafer by using a (200)-oriented TiN film as a buffer layer [14, 15].

While the fabricated NbN/AlN/NbN epitaxial junctions have shown promising results, challenges remain in developing a reliable fabrication process for sub-micron or deep-sub-micron junctions as required in large scale superconducting qubit circuit. We have developed a new fabrication process for deep-sub-micron Josephson junctions by using an electron beam (EB) lithography and planarization process with chemical mechanical polishing (CMP), as described in the following section. Full-epitaxial NbN/AlN/NbN junctions with low leakage current were successfully fabricated with a junction size as small as 0.27 µm in diameter. After measuring the current-voltage (*I-V*) characteristics of the junction, the SiO$_2$ dielectric layer, which served as the isolation between the base electrode and the wiring layer, was removed in a buffered HF solution. The *I-V* characteristics were re-measured and compared to the original properties.

2. Fabrication Process

2.1 Preparation of full-epitaxial NbN/AlN/NbN tri-layer

Details of the fabrication process of full-epitaxial NbN/AlN/NbN tri-layer structure on a Si (100) wafer can be found in our early report [14, 15]. Firstly, a 40-nm-thick TiN film was deposited on a hydrogen-terminated Si substrate at a substrate temperature of 850°C with a total outgas pressure of $1\times10^{-6}$ Pa by dc magnetron sputtering method. The background pressure without substrate heating was better than $8\times10^{-8}$ Pa. X-ray diffraction (XRD) analysis revealed a single (200)-oriented reflection peak at 42.66° in the $\theta/2\theta$ scan, resulting in a lattice constant $a_0$=0.423 nm. The TiN (200) film exhibited a superconducting transition temperature $T_c$ of 5.6 K, a resistivity at 10 K of 3.2 μΩ cm, and a residual resistivity ratio (the ratio of resistivities at 300 K and 10 K) of 5.4. The lattice constant of TiN (200) film is therefore relatively close to that of a (200)-oriented NbN film deposited at optimal conditions ($a_0$=0.446 nm). The thin TiN film serves as a buffer layer for the epitaxial growth of NbN film, as well as for the NbN/AlN/NbN tri-layer.

After the initial growth of (200)-oriented TiN film, the wafer was cooled down below 100°C before the NbN/AlN/NbN tri-layer deposition in-situ by dc reactive sputtering without breaking vacuum. The 100 nm thick base and 200 nm thick counter NbN layers were deposited in the mixture of Ar and $N_2$ gases, while a 2 nm thick AlN tunnel barrier was deposited in a pure $N_2$ gas. The $T_c$ of the 200-nm thick NbN film on the TiN buffer was 15.3 K, which is almost equal to that of on a MgO (100) substrate [13]. We examined the crystal structure of NbN/AlN/NbN tri-layer by XRD analysis and could not identify any other peaks apart from a single (200) peak, which suggests the growth of a cubic structure of AlN barrier with a thickness up to 2 nm. Additional information regarding the AlN thickness dependence of the crystal structure of the counter NbN electrode, as well as the crystal structure of the AlN tunnel barrier, can be found in our early work [9, 11].

2.2 Fabrication of small Josephson junctions with a planarization process

Figure 1 illustrates the fabrication process of small Josephson junctions with an electron beam lithography (EBL) and a planarization process. Firstly, the base electrodes of the junction were patterned by an i-line stepper (Canon FPA-3030 i5+) using a positive photoresist AZ-MiR 703 (Integrated Micro Material Inc.). The NbN/AlN/NbN tri-layer and the TiN buffer layer was etched by a reactive ion etching (RIE) using $CF_4$ gas for NbN and TiN, and Ar gas for AlN. After removing the photoresist, the junctions were patterned by EBL using a ZEP-520A positive resist with a thickness of 320 nm, where an ELIONIX ELS-F125 system with an electron beam of 125 kV was employed. A thin Al or MgO layer was deposited by dc sputtering, followed by a lifted-off in NMP solution. This thin layer served as an etching mask for the junction definition in RIE with a $CF_4$ plasma. The mask layer was eliminated by a wet or dry etching subsequently after the junction RIE process.

The $SiO_2$ was deposited by rf sputtering to provide the interlayer isolation between the base electrodes and the wiring layer. Here, the thickness of $SiO_2$ was determined from the thickness of the tri-layer, including the TiN buffer layer, the over-etched Si, and an additional 100 nm for the following planarization process by CMP. A total thickness of 500 nm $SiO_2$ was deposited. To avoid the pattern size dependence of the polishing rate in the CMP process, we employed a caldera-based planarization that has been developed for the SFQ circuitry process [17, 18]. Caldera walls with a width of 1.5 μm were created at the edges of base electrodes by RIE. By etching the $SiO_2$ for an appropriate time, the surface of $SiO_2$ on the base electrode is approximately at the same height as the surface of $SiO_2$ on the Si wafer, as shown in figure 1(e). Then, the wafer was planarized by CMP in a TRCP380 CMP system (Techno Rize Inc.). As a result, as illustrated in figure 1(f), the wafer surface can be completely planarized by the CMP process with a relatively short polishing time, with no dependency on the pattern size.

The contact via-holes were patterned by an i-line stepper followed by an RIE to a depth of 200 nm, which equals to the height of the counter electrode. The remaining $SiO_2$ was etched by a $CHF_3$ plasma until the depth of contact via-hole is less than 200 nm. The AlN/NbN acted as an etching stopper for the $CHF_3$ plasma. Multiple etches in the $CHF_3$ plasma of short duration were needed to prevent over-etching of the $SiO_2$. The step profiler verified the etching progress after each step. The caldera-based planarization and $CHF_3$ etching process provide a reliable solution to form the junction and the base contacts without having to measure the thickness of $SiO_2$ with an ellipsometer.

Finally, a NbTiN wiring layer with a thickness of 300 nm was deposited and patterned. We measured the *I-V* characteristics of the junction in a liquid helium environment. Unwanted TLSs in the amorphous $SiO_2$ can contribute towards a significant dielectric loss at low microwave powers and low temperatures. To remove the $SiO_2$ dielectric layer, we immersed the sample in a buffered HF (BHF) for 10 minutes. Figure 2 shows the scanning electron microscope (SEM) images of the fabricated junctions with the design junction sizes of 0.7 μm and 0.4 µm in diameters after removing the SiO$_2$ interlayer. Both junctions are somewhat smaller than the designed sizes because of the shrinkage during the junction RIE process. The estimation of the actual junction size from its electrical properties will be described in the next section.

3. Electrical properties of NbN/AlN/NbN junctions

3.1 Junction properties before and after SiO$_2$ removal

To evaluate the junction properties, we measured the *I-V* characteristics of the fabricated junctions at 4.2 K. Figure 3 shows a typical *I-V* curve of the NbN/AlN/NbN epitaxial junction with a designed junction size of 0.9 µm in diameter. The black curve and the red curve are the measured *I-V* curves before and after removing the SiO$_2$ dielectrics layer, respectively. Both curves show a similar trend, indicating that the wet etching process by BHF did not have a significant impact on the properties of the junction. The normal resistance of the junction, $R_N$, measured at 10 mV, is approximately 24.9 kΩ. The sub-gap resistance, $R_{SG}$, estimated at 3 mV, is 3 MΩ. The ratio of $R_{SG}/R_N$ is 121, which suggests an excellent quality of the tunnel barrier with a very low leak current.

3.2 Size dependence of junction properties

The novel junction fabrication process using EBL and planarization offers a high device yield for all junctions with designed junction sizes from 20×20 µm$^2$ to as small as 0.4 µm in diameter. Figure 4 shows *I-V* characteristics after removing the SiO$_2$ for the fabricated junctions with a variety of junction sizes. All junctions show a clear gap structure. The critical current density, $J_c$, of the junction is defined as $I_c/A$, where $I_c$ is the critical current of tunnel junction, and $A$ is the junction area. However, due to thermal or environmental noise in our measurement system, direct measuring $I_c$ for values less than 1 µA was difficult. Therefore, we extracted $I_c$ from the Ambegaokar-Baratoff relation, $I_g\pi/4$, where $I_g$ is the quasiparticle current measured at the gap voltage [19].

In figure 4, the $R_NA$ product and $J_c$ have been calculated for junctions with different sizes. Both the $R_NA$ product and $J_c$ should be constant regardless of the junction size [20]. However, the result in figure 4 indicates that the $R_NA$ product increases, and $J_c$ decreases, as the junction size decreases. This suggests that the actual junction areas are smaller than the designed values because of shrinkage from over-etching the junction during RIE, rather than the high accurate EBL.

In figure 5, we plot the junction area dependence of $R_NA$ assuming various junction shrinkage values, $\Delta d$. If zero shrinkage is assumed ($\Delta d=0$ μm, the black curve in figure 5), there is an inverse proportionality between the junction size and the $R_NA$ value. This result coincides with the results presented in figure 4. Conversely, a $\Delta d$ of -0.25 μm resulted in a direct proportionality between the junction size and the $R_NA$ value, as shown in the blue curve in figure 5. If we assume a $\Delta d$ of -0.13 μm, the $R_NA$ product becomes virtually constant regardless of the junction area, as indicated by the red curve in figure 5. The designed junction diameters of 0.7 μm and 0.4 μm in figure 2 are reduced to 0.57 μm and 0.27 μm, which seems to be consistent with the SEM observations. By taking into account a $\Delta d$ of -0.13 μm, it is possible to re-calculate $J_c$ values for the junctions shown in figure 4, resulting in a range of 57.6-61.6 kA/cm$^2$.

3.3 Global variation of junction properties

We verified the global variation of $J_c$ and $\Delta d$. Figure 6 summaries the measured results of six chips (2×5 mm$^2$) with their relative locations within CHIP-A to CHIP-F, representing different areas across the 2in-wafer. The junction shrinkage $\Delta d$ of each measured chip was extracted similar to that of the junction area independence of $R_NA$, as shown in figure 5. The extracted $\Delta d$ is virtually constant at -0.13 μm for CHIP-A to CHIP-E. CHIP-F showed a slightly larger $\Delta d$ value of -0.15 μm. The slightly larger $\Delta d$ is considered within a reasonable error margin due to the non-uniformity in the RIE etching rate across the wafer, being slightly higher at its periphery than in its center.

Similarly, we extracted the $J_c$ distribution within a range from 40 to 59 A/cm$^2$ across the wafer. We attribute the variation of $J_c$ to the non-uniformity of the AlN tunnel barrier. The AlN tunnel barrier is deposited by dc reactive sputtering with an 8-inch magnetron cathode. However, a non-uniform magnetic field on the cathode may lead to a thickness distribution of the AlN tunnel barrier. For future work, a rotational substrate holder for the deposition may address this concern.

4. Conclusion

We have developed a novel fabrication process for deep-sub-micrometer NbN/AlN/NbN epitaxial junctions by EBL and planarization with CMP on a Si substrate. The fabricated junctions showed a clear gap structure with designed junction size as small as 0.4 μm in diameter. All junctions have very high sub-gap resistance, $R_{SG}$, with an $R_{SG}/R_N$ ratio over 100, suggesting small leakage currents and a high-quality AlN tunnel barrier. After the initial measurement of the *I-V* curves, the SiO$_2$ dielectric layer used as the isolation between the base and the wiring layers was removed by wet etching in a BHF solution. By comparing the *I-V* curves before and after the removal of SiO$_2$, we found that wet etching with BHF has no significant impact on the junction properties. The junction shrinkage, $\Delta d$, was estimated to be 0.13 μm from the junction area independence of $R_N A$ product, leading to an actual junction diameter of 0.27 μm for the design size of 0.4 μm. The $\Delta d$ across the 2-inch wafer is nearly constant at 0.13 μm. A slightly larger shrinkage $\Delta d$ was found at the periphery of the wafer due to the differences in the etching rate of NbN during the junction definition. The $J_C$ was extracted from chips at different locations and ranged from 40 to 59 A/cm$^2$, suggesting a non-uniform AlN tunnel barrier across the wafer. Our newly developed fabrication process for deep-sub-micrometer NbN/AlN/NbN epitaxial junctions can be applied for a wide range of superconducting electronic applications, such as superconducting qubits, superconducting quantum interference devices (SQUIDs), SFQ logic circuits.

Acknowledgments

The authors would like to thank Dr. Akira Kawakami for fruitful discussion and Dr. Saburo Imamura for the technical support. This work was supported by the Japan Science and Technology Agency under ERATO (JPMJER1601) and CREST (JPMJCR1775).

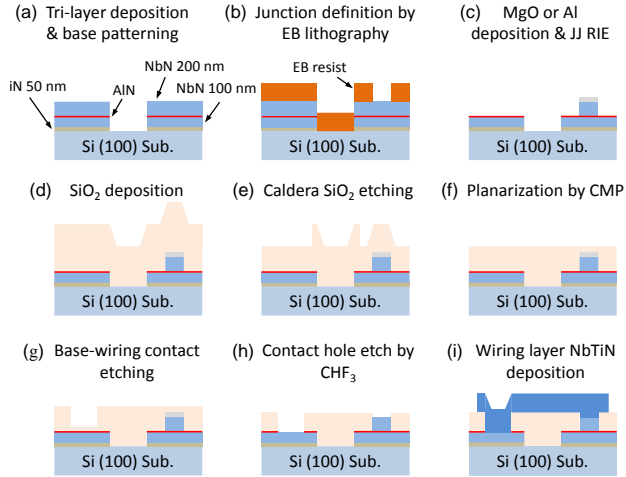

**Figure 1.** Schematic fabrication process flow diagram for ultra-small epitaxial NbN tunnel junctions by EB and CMP processes: (a) Tri-layer NbN/AlN/NbN deposition and base patterning definition; (b) Junction definition by EB lithography; (c) Formation of Al etching mask for junction definition by RIE; (d) Deposition of interlayer dielectrics $SiO_2$; (e) Formation of caldera patterns by dry etching $SiO_2$ in $CF_4$ plasma; (f) Wafer polishing by CMP; (g) Definition of contact hole between base and wiring layer; (h) Contact hole etch back by $CHF_3$ plasma; (i) Formation of NbTiN wiring layer; Final step (not shown): Removal of $SiO_2$ dielectrics by diluted HF solution.

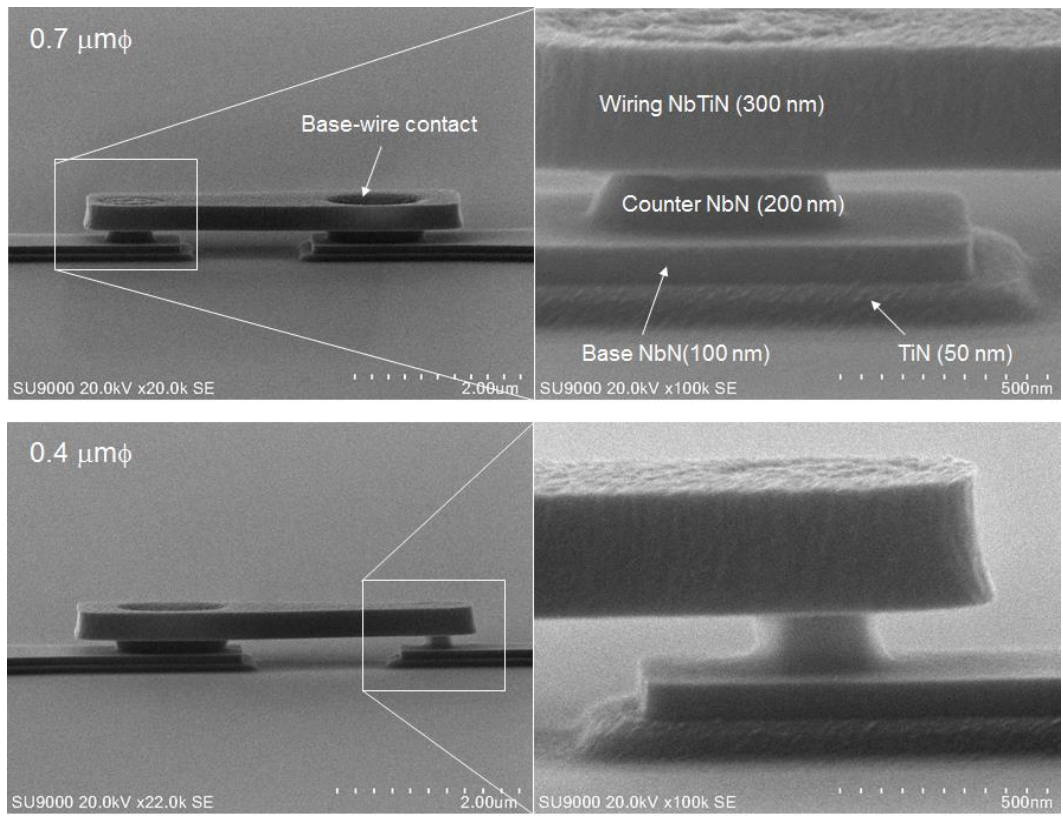

**Figure 2.** SEM images of NbN/AlN/NbN epitaxial junctions with the designed sizes of 0.7 μm and 0.4 μm in diameters. The $SiO_2$ used as the isolation between the base and the wiring layers has been eliminated by a wet etching with BHF.

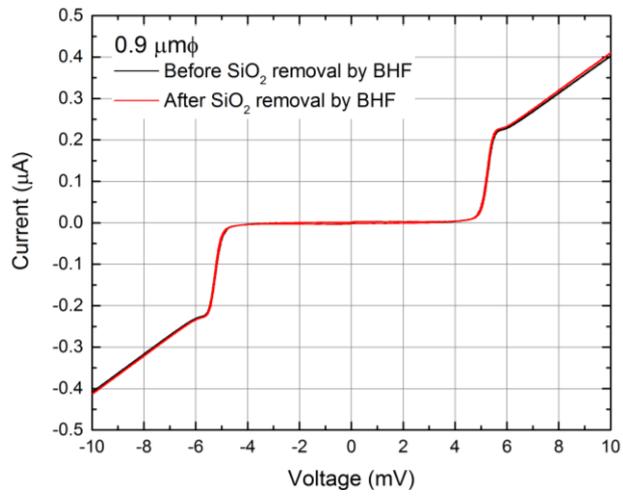

**Figure 3.** *I-V* characteristics of NbN/AlN/NbN epitaxial tunnel junction with a designed junction size of 0.9 μmϕ measured at 4.2 K. Black and red colored curves represent the data before and after removing $SiO_2$ dielectric layer by BHF, respectively.

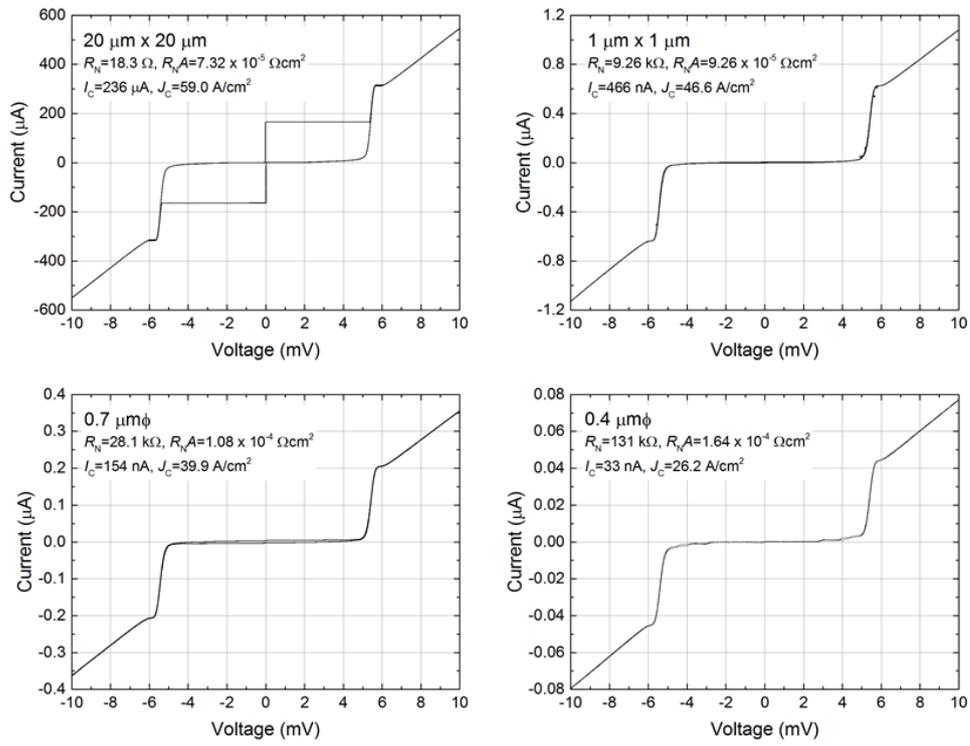

**Figure 4.** *I-V* characteristics of NbN/AlN/NbN epitaxial junctions with various junction sizes ranged from 20 μm in square to 0.4 μm in circular. The displayed junction sizes are not the actual size, but the designed size.

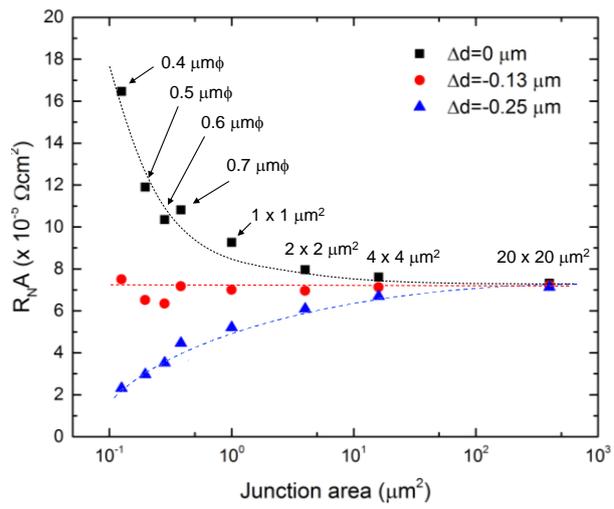

**Figure 5.** Junction area dependence of $R_NA$ for various junction shrinkages, $\Delta d$, due to the junction RIE process: 0 μm (black curve); -0.13 μm (red curve); -0.25 μm (blue curve).

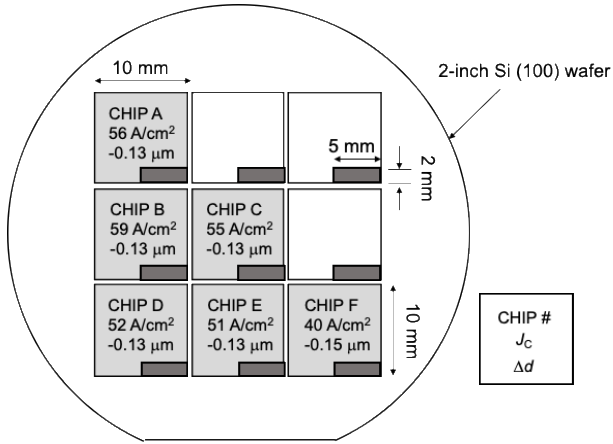

**Figure 6.** Global variation of $J_c$ and $\Delta d$ across 2-inch wafer.